\documentclass[11pt,a4paper]{article}
\usepackage{amsmath,amssymb}
\usepackage{epsfig,graphicx}

\begin{document}

\title{\bf Three-loop SM RGEs with general Yukawa matrices}
\author{A.~V.~Bednyakov$^a$\footnote{{\bf e-mail}: bednya@theor.jinr.ru}
\\
$^a$ \small{\em Joint Institute for Nuclear Research} \\
\small{\em 141980 Dubna, Russia}
}
\date{}
\maketitle

\begin{abstract}
The results for the three-loop renormalization group equations 
for all fundamental parameters of the SM Lagrangian are presented. Special attention is paid to the
Flavor sector of the SM, which parameterized by general complex non-diagonal Yukawa couplings.
Some details of calculation techniques are given. 
In addition, ambiguities in the beta-functions for the matrix couplings are discussed. 
\end{abstract}

The discovery of the Higgs boson at the LHC \cite{Chatrchyan:2012ufa,Aad:2012tfa} leads to a revival of interest to 
the analysis of the high-energy behavior of the SM couplings (see, e.g., Refs.~\cite{Bezrukov:2012sa,Degrassi:2012ry}) 
	based on renormalization group  equations (RGEs).
During past two years three-loop contributions to RGEs of all SM parameters were 
obtained~\cite{Mihaila:2012fm,Mihaila:2012pz,Chetyrkin:2012rz,Bednyakov:2012rb,Bednyakov:2012en,Chetyrkin:2013wya,Bednyakov:2013eba}
	extending well-known two-loop results (see, e.g., \cite{Luo:2002ey} and references therein).
However, in most of the mentioned three-loop calculations the flavor structure of the SM was simplified and only diagonal 
	 fermion couplings to the Higgs boson were considered. 
The only exception was the pioneering calculation of Ref.~\cite{Mihaila:2012pz}, in which the three-loop gauge-coupling beta-functions 
	with the account of matrix Yukawa interactions were recovered from the expressions obtained in the flavor-diagonal SM.

In a series of papers \cite{Bednyakov:2013cpa,Bednyakov:2014pia} our group 
	re-calculated three-loop RGEs for the SM couplings taking into account explicit flavor indices, thus,
	allowing one to study the origin of the SM Flavor pattern, which could originate from some New Physics.

To set notation, let us define the following quantities for the dimensionless parameters of the SM:
\begin{equation}
	 a_i   =   h \cdot \left(\frac{5}{3} g_1^2, g_2^2, g_s^2\right), 
	\quad \hat \lambda =   h \cdot \lambda, \qquad h \equiv \frac{1}{16\pi^2}.
	\label{eq:coupl_notation}
\end{equation}
	Here  $g_s$, $g_2$, $g_1$ correspond to the running $SU(3)\times SU(2) \times U(1)$ gauge couplings,
	and $\lambda$ --- to the Higgs self-coupling in the unbroken SM (see, e.g., \cite{Bednyakov:2012rb}).
In addition, the following abbreviations ($f,~f' = u, d, l$)
\begin{equation}
	\mathcal{Y}_{\underbrace{ff'\dots}_n} \equiv h^n \left(Y_f Y_f^\dagger Y_f' Y_f'^\dagger\dots\right),
	\label{eq:matrix_notation}
\end{equation}
	will be used to denote different Yukawa matrix products.

	In Eq.\eqref{eq:matrix_notation} the matrix element $Y^{ij}_f$ describes the transition 
  of the right-handed fermion $f$ of the $j$-th generation to the left-handed fermion (either up- or down-type) 
  of the generation $i$. 
Conversely, the matrix element $Y^{\dagger,ij}_f$ corresponds to the transition of some left-handed fermion from an
  SU(2) doublet of the generation $j$ to the right-handed SU(2) singlet $f_i$. 

It is worth noting that not all of the matrix elements are ``observable'' quantities. 
This is due to the fact that for any given set of Yukawa matrices one can use an accidental global symmetry of
	the SM gauge interactions 
	$U(3)_Q\times U(3)_u \times U(3)_d \times U(3)_L \times U(3)_l$ to 
	choose a specific basis in flavor space, in which Yukawa sector of the SM can be parameterized only
	by 13 parameters. 
The latter correspond to the nine fermion masses\footnote{We neglect neutrino masses and mixing.} 
	(related via spontaneous symmetry breaking to the Yukawa couplings),
	three mixing angles and one CP-violating phase of the Cabibbo-Kobayashi-Maskawa (CKM) matrix $V_{\mathrm{CKM}}$.
The CKM matrix is a product of unitary factors $U_L$ and $D_L$ that diagonalize quark Yukawa matrices (mass basis)
\begin{eqnarray}
	U_L Y_u U^\dagger_R = Y^u_{\mathrm{diag}},\qquad D_L Y_d D_R^\dagger = Y^d_{\mathrm{diag}}, \qquad V_{\mathrm{CKM}} \equiv U_L D^\dagger_L.	
	\label{eq:yuk_diag}
\end{eqnarray}
	Note that the above-mentioned global symmetry imply equivalence relations  
\begin{eqnarray}
	&&\left( Y_u, Y_d, Y_l \right) 
 \Leftrightarrow 
	\left( Y'_u, Y'_d, Y'_l \right) = 
	\left( V_Q Y_u V_u^\dagger, 
	       V_Q Y_d V_d^\dagger,  
	       V_L Y_l V_l^\dagger  
	\right),
	\qquad
	V_f \in U(3)_f, 
	\label{eq:equiv_classes}
\end{eqnarray}
	so that both $Y_q$ and $Y'_q$ lead to the same  $Y^q_{\mathrm{diag}}$ and $V_{\mathrm{CKM}}$ with, 
	e.g., $U_L' = U_L V_Q^\dagger$, $D_L' = D_L V_Q^\dagger$, etc.  
This observation will be important in the following discussion of the Yukawa matrix beta-functions and their ambiguities.

Before going to the results, let us briefly mention the utilized techniques and methods. 
In order to carry out such a calculation we made use of different computer codes, both private and public ones. 
Due to the fact that RGEs
	for fundamental parameters of the SM Lagrangian (i.e., dimensionless couplings together with the Higgs potential mass parameter)
	can be found solely from ultraviolet asymptotics of different Green functions,
	one can perform the calculation within the unbroken phase of the model with massless fields.  
The beta-functions for the SM parameters are extracted from the corresponding {\ensuremath{\overline{\mathrm{MS}}}} renormalization constants involving only poles in $\epsilon = (4-d)/2$.

The required Feynman rules of the ``unbroken'' SM  with matrix Yukawa couplings were automatically obtained from the SM Lagrangian 
	by means of {\tt LanHep} package \cite{Semenov:2010qt}.
The output of {\tt LanHep} was utilized by {\tt FeynArts} \cite{Hahn:2000kx} and, independently, by {\tt DIANA} \cite{Tentyukov:1999is}
	to generate Feynman diagrams and the corresponding expressions. 
	It is due to nice features of dimensional regularization and {\ensuremath{\overline{\mathrm{MS}}}} scheme one can further simplify the calculation procedure. 
Since we are interested only in ultraviolet (UV) behavior of Green-functions the infrared (IR) structure  
	of the corresponding Feynman integrals can be modified \cite{Vladimirov:1979zm} (so called infrared rearrangement or IRR)  to convert all the relevant diagrams either to massless self-energies
	or to fully massive vacuum graphs.
The former are evaluated by means of {\tt MINCER} \cite{Gorishnii:1989gt,Larin:1991fz} and the latter can be computed 
	either with the public \texttt{MATAD} package \cite{Steinhauser:2000ry} or private \texttt{BABMA} code written by Velizhanin.

In order to obtain the presented results we made use of both options.
The reason is quite obvious: such a complicated calculation requires evaluation of several millions\footnote{This number can be signifcantly reduced by means of the {\tt GraphState} library \cite{Batkovich:2014bla}.}  of Feynman diagrams and can 
	only be carried out by modern computers. 
Due to this, the results should be thoroughly cross-checked prior to publication. 
For example, the renormalization constants used to derive the SM RGEs should be free from gauge-fixing parameter dependence and satisfy the
	pole equations \cite{'tHooft:1972fi}.

The evaluation of the gauge-coupling beta-functions and that of the parameters of the Higgs potential did not pose significant problems, since
	one can use existing codes modified to handle explicit flavor indices. 
Consequently, we will not spend much time discussing the results given in Ref.~\cite{Mihaila:2012pz,Bednyakov:2012rb,Bednyakov:2013cpa} and concentrate
	on the beta-functions for the matrix couplings.
The only thing we would like to mention is that the full three-loop results for $\lambda$ and the Higgs mass parameter $m^2$ 
	can not be recovered be means of substitutions \cite{Mihaila:2012pz} like
\begin{eqnarray}
	n_Y^2 y_u^2 y_d^2 \to \mathrm{tr} \left( \mathcal{Y}_u \right) \mathrm{tr}\left( \mathcal{Y}_d \right), \qquad
	n_Y y_u^2 y_d^2 \to \mathrm{tr} \left(\mathcal{Y}_u \mathcal{Y}_d \right)
	\label{eq:tricks}
\end{eqnarray}
	with $n_Y$ counting  fermion loops with at least two couplings to the Higgs field and $y_u$, $y_d$ being diagonal Yukawa couplings.  
It is obvious that Yukawa interactions contribute to the running of $\lambda$ and $m^2$ starting from the one-loop level (contrary to the running of the gauge couplings)
so one is forced to treat flavor explicitly to, e.g., distinguish between contributions of the 
	form $\mathrm{tr}\left(\mathcal{Y}_f\right) \mathrm{tr}\left(\mathcal{Y}_f\mathcal{Y}_f\mathcal{Y}_f\right)$ and  
	$\mathrm{tr}\left(\mathcal{Y}_f \mathcal{Y}_f\right) \mathrm{tr}\left(\mathcal{Y}_f\mathcal{Y}_f\right)$.	

The matrix coupling renormalization constants needed to find the corresponding beta-functions were obtained for the first time by means
	of the procedure based on {\tt MINCER} code (see, e.g., Ref.~\cite{Bednyakov:2013cta} for details), which only requires evaluation of bare Green functions
	without explicit counter-term insertions. 
However, albeit the fact that the renormalization constants $\Delta Y_f$ for $Y_f$, which relate bare and renormalized couplings ($\mu$ - renormalization scale)
\begin{equation}
	\left(Y_f\right)_{\mathrm{Bare}}  =  \left(Y_f + \Delta Y_f \right) \mu^{\epsilon}, 
	\label{eq:Y_bare_to_ren}
\end{equation}
	were free from gauge-parameter dependence the corresponding beta-functions defined by
\begin{equation}
	\beta_{Y_f} Y_f \equiv \frac{d Y_f(\mu,\epsilon)}{d \ln \mu^2}\bigg|_{\epsilon=0} 
	\label{eq:Yuk_beta_def}
\end{equation}
	were divergent in the limit $\epsilon \to 0$.
In other words, the pole equations that guarantee  the absence of negative powers of $\epsilon$ in the relation 
\begin{eqnarray}
	\beta_{Y_f} Y_f = - \left( \frac{d}{d\ln \mu^2} + \frac{\epsilon}{2} \right) \Delta Y_f	
	 \label{eq:Yuk_beta_deriv}
\end{eqnarray}
	were not satisfied.
We re-evaluated all the required Green functions by means of {\tt MATAD / BAMBA} setup \cite{Bednyakov:2013cta}, 
	which employ different approach \cite{Misiak:1994zw,Chetyrkin:1997fm,Zoller:2014xoa} to IRR.
This independent calculation confirms that there was no error in counter-terms for Green-functions in the first attempt. 
However, the procedure used to extract matrix renormalization constants for Lagrangian parameters contains a freedom, 
	which can be utilized to resolve the encountered issue.

Essentially, one can not deduce unambiguously the renormalization matrix for the quark doublet $Q = (u_L,d_L)$ and singlet $(u_R,d_R)$ fields
	defined as
\begin{eqnarray}
	Q_{L,\mathrm{Bare}}  =  \left[Z^{1/2}_{Q}\right] Q_L \mu^{-\epsilon}, \quad  
	&u_{R,\mathrm{Bare}}  =  \left[Z^{1/2}_{u}\right] u_R \mu^{-\epsilon},& 
	\quad d_{R,\mathrm{Bare}}  =  \left[Z^{1/2}_{d}\right] d_R \mu^{-\epsilon}, 
	\label{eq:quarks_ren}
\end{eqnarray}
	from a self-energy counter-term
\begin{equation}
	i \hat p P_L \left[\left(Z^{1/2,\dagger}_{f_L} Z^{1/2}_{f_L} \right)_{ij} - \delta_{ij}\right] + (L \to R), \quad P_{L,R} = \frac{1\mp\gamma_5}{2}.
	\label{eq:se_ct}
\end{equation}
	This ``square root'' ambiguity allows us to introduce additional unitary factors  
\begin{eqnarray}
	\bar{Z}^{1/2}_{Q_L} & = & 
		1  -  
		 \frac{a_1}{320} \left(\frac{1}{6 \epsilon^2} - \frac{1}{\epsilon^3} 
		 \right) 
		 \bigg[\mathcal{Y}_u,\mathcal{Y}_d\bigg]
		+ \frac{1}{64} \left(\frac{1}{6 \epsilon^2} + \frac{1}{\epsilon^3} \right) 
		\bigg\{ \mathcal{Y}_u - \mathcal{Y}_d, \bigg[ \mathcal{Y}_u ,\mathcal{Y}_d \bigg] \bigg\}
		,	 
	\nonumber\\
	\bar{Z}^{1/2}_{f_R} & = & 
	1  \mp \frac{1}{32}  \left(\frac{1}{6 \epsilon^2} - \frac{1}{\epsilon^3} \right) 
		Y^\dagger_f \bigg[
		\mathcal{Y}_u ,\mathcal{Y}_d 
	\bigg] Y_f,\qquad \mathrm{for}~f = u,~d	
	\label{eq:ren_ferm_nonherm}
\end{eqnarray}
	where the commutator $\left[\mathcal{Y}_u ,\mathcal{Y}_d \right]$ is an 
	anti-hermitian matrix, which is a measure of whether $Y_u$ and $Y_d$ can be diagonalized simultaneously.
The factors \eqref{eq:ren_ferm_nonherm} are combined with the hermitian ``square roots'' $\tilde Z^{1/2}_f$ 
	extracted in perturbation theory from \eqref{eq:se_ct} to give 
	\begin{equation}
Z_f^{1/2} = \bar Z_f^{1/2} \tilde Z_f^{1/2},~
	\bar Z_f^{1/2\dagger} = \bar Z_f^{-1/2}, ~ \tilde Z_f^{1/2\dagger} = \tilde Z_f^{1/2}.
		\label{eq:full_quark_rc}
	\end{equation}
The latter leads to our final (and finite in the limit $\epsilon \to 0$) result 
	for the quark field anomalous dimensions 
\begin{eqnarray}
	\gamma_{f} = - \left(Z_f^{-1/2} \frac{d}{d\ln\mu^2} Z_f^{1/2} \right),~
	\label{eq:fermion_anom_dim_def}
\end{eqnarray}
and the Yukawa matrix beta-functions extracted via \eqref{eq:Yuk_beta_deriv} from 
\begin{equation}
	Y_f + \Delta Y_f =   
	\left[Z_{f'}^{-1/2}\right]^\dagger Z_{\bar{f}'f\phi} Y_f \left[Z_f^{-1/2}\right] Z_\phi^{-1/2},
	\label{eq:yuk_ren}
\end{equation}
	with $(Z_{\bar{f}'f\phi} - 1) Y_f$ being the fermion-fermion-Higgs $ff'\phi$ vertex counter-term and $Z_{\phi}^{1/2}$
	corresponding to the Higgs fields renormalization constant. 

We would like to emphasize that different $\bar Z_f^{1/2}$ from \eqref{eq:ren_ferm_nonherm} 
	do not modify first poles in $\epsilon$ of the resulting $Z_f^{1/2}$. 
However, it is possible to introduce additional unitary factors, which do involve the $1/\epsilon$ terms  
	(see \cite{Bednyakov:2014pia} for explicit example)
	and, as a consequence, lead to a modification of the matrix quark anomalous dimensions and Yukawa beta-functions 
	already at the two-loop level. 
Nevertheless, all these unitary factors are ``non-physical'', since they relate Yukawa matrices belonging to the same equivalence classes
\eqref{eq:equiv_classes}.
Due to this, the renormalization group flow of the observable parameters 
(e.g., $Y^f_{\mathrm{diag}}$ or CKM parameters) does not affected.
Some extended discussion on this topic together with explicit results can be found in our recent publication~\cite{Bednyakov:2014pia}.

To conclude, three-loop beta-functions for all SM Lagrangian parameters were calculated with the account of matrix Yukawa couplings.
The obtained expressions can be applied to RGE studies of different New Physics models aimed to unveil the dynamics
	behind the observed SM flavour pattern.
It is also worth mentioning that from $Y_u$ and $Y_d$ it is possible 
to deduce the three-loop RGEs for the CKM matrix elements \cite{Babu:1987im,Naculich:1993ah,Balzereit:1998id} 
or Quark Flavour invariants~(see, e.g., Ref.~\cite{Jenkins:2009dy}).

\section*{Acknowledgments}
	The author thanks A.F.~Pikelner and V.N.~Velizhanin for a fruitful collaboration on the topic.
	Special thanks go to the Organizing committee of the QUARKS-2014 seminar for warm hospitality.
In addition, the speaker is grateful to A.Pivovarov for stimulating discussions during the conference. 
This work is supported in part by RFBR grant 12-02-00412-a, 
	the Ministry of Education and Science of the Russian Federation Grant MK-1001.2014.2,
	and the JINR grant 14-302-02.


\end{document}